\documentclass[useAMS,usenatbib,usegraphicx]{mn2e}

\title[A direct $N$-body model]{A direct $N$-body model of core-collapse and core oscillations}  
\author[J. R. Hurley \& M. M. Shara]{Jarrod R. Hurley$^{1}$\thanks{
E-mail: jhurley@swin.edu.au (JRH)}, 
Michael M. Shara$^2$ \\
$^{1}$Centre for Astrophysics and Supercomputing, Swinburne University of Technology, P.O. Box 218, VIC 3122, Australia \\
$^{2}$Department of Astrophysics, American Museum of Natural History, \\
Central Park West at 79th Street, New York, NY 10024, USA}

\begin{document}

\date{Received 2012 Month xx; in original form 2012 March 15} 

\pagerange{\pageref{firstpage}--\pageref{lastpage}} \pubyear{2012}

\maketitle

\label{firstpage}

\begin{abstract}
We report on the results of a direct $N$-body simulation of a star cluster that started with
$N = 200\,000$,
comprising $195\,000$ single stars and $5\,000$ primordial binaries.
The code used for the simulation includes stellar evolution, binary evolution, an external
tidal field and the effects of two-body relaxation.
The model cluster is evolved to $12\,$Gyr, losing more than 80\% of its stars in the process.
It reaches the end of the main core-collapse phase at $10.5\,$Gyr and experiences core
oscillations from that point onwards 
-- direct numerical confirmation of this phenomenon. 
However, we find that after a further $1\,$Gyr the core oscillations are halted by the 
ejection of a massive binary comprised of two black holes from the core, 
producing a core that shows no signature of the prior core-collapse. 
We also show that the results of previous studies with $N$ ranging from
$500$ to $100\,000$ scale well to this new model with larger $N$. 
In particular, 
the timescale to core-collapse (in units of the relaxation timescale), 
mass segregation, velocity dispersion, 
and the energies of the binary population 
all show similar behaviour at different $N$. 
\end{abstract}


\section{Introduction}
\label{s:intro}

The increasing ability of the direct $N$-body method to provide reliable
models of the dynamical evolution of star clusters has closely mirrored
increases in computing power (Heggie 2011).
The community has progressed from small-$N$ models performed on workstations 
(e.g. von Hoerner 1963; McMillan, Hut \& Makino 1990; Giersz \& Heggie 1997) 
to models of old open clusters 
and into the $N \sim 100\,000$ regime (Baumgardt \& Makino 2003) 
by making use of special-purpose GRAPE hardware (Makino 2002). 
Software advances over a similar timeframe have produced sophisticated
codes (Aarseth 1999; Portegies Zwart et al. 2001) that increase the realism of the
models by incorporating stellar and binary evolution, binary formation, three-body
effects and external potentials. 
As a result, $N$-body models have been used in numerous ways to understand
the evolution of globular clusters
(GCs: Vesperini \& Heggie 1997; Baumgardt \& Makino 2003; Zonoozi et al. 2011),
even though the best models still only
touch the lower end of the GC mass-function 
(see Aarseth 2003 and Heggie \& Hut 2003 for a more detailed review of previous work). 
At the other end of the spectrum, Monte Carlo (MC) models have proven effective
at producing dynamical models of $N > 10^6\,$particles (Giersz \& Heggie 2011). 
These models have shown that 
clusters previously defined as 
non-core-collapse can actually be in a fluctuating post-core-collapse phase 
(Heggie \& Giersz 2008). 
In practice the two methods are complimentary with MC informing the more
laborious $N$-body approach (such as refining initial conditions) and
$N$-body calibrating aspects of MC.

In this paper we present an $N$-body simulation of star cluster evolution that
begins with $N = 200\,000$ stars and binaries.
This extends the $N$ parameter space covered by direct $N$-body models and performs
two important functions.
Firstly it provides a new calibration point for the MC method
-- this statistical method is increasingly valid for increasing $N$ so calibrations
at higher $N$ are more reliable.
It also allows us to further develop our theoretical understanding of star
cluster evolution and investigate how well inferences drawn from models of smaller $N$
scale to larger values. 
The latter is the focus of this current paper. 
A good example of the small-$N$ models that we wish to compare with is the 
comprehensive study of star cluster evolution presented by Giersz \& Heggie (1997) 
using models that included 
a mass function, stellar evolution and the tidal field of a point-mass galaxy, albeit
starting with $500$ stars instead of $200\,000$. 
More recent examples for comparison include Baumgardt \& Makino (2003) 
and K\"{u}pper et al. (2008). 
We were also motivated to produce a model that exhibited core-collapse 
close to a Hubble time without dissolving by that time. 
What we find when interpreting this model is that much of the behaviour reported 
previously for smaller $N$-body models stands up well in comparison 
but that the actions of a binary comprised of two black holes (BHs) provides a late twist 
to the evolution of the cluster core. 

In Section 2 we describe the setup of the model. 
This is followed by a presentation of the results in Sections 3 to 7 focussing on 
general evolution (cluster mass and structure), the impact of the BH-BH binary, 
mass segregation, velocity distributions and binaries (binary fraction and binding energies). 
Throughout these sections the results are discussed and compared to previous 
work where applicable. 
Then in Section 8 we specifically look at how the evolution timescale of the new 
model compares to findings presented in the past.

\section{The Model}
\label{s:models}

For our simulation we used 
the {\tt NBODY4} code (Aarseth 1999)
on a GRAPE-6 board (Makino 2002)
located at the American Museum of Natural History.
{\tt NBODY4} uses the 4th-order Hermite integration scheme
and an individual timestep algorithm to follow the orbits of cluster
members and invokes regularization schemes to deal with the
internal evolution of small-$N$ subsystems
(see Aarseth 2003 for details).
Stellar and binary evolution of the cluster stars are performed in
concert with the dynamical integration as described in Hurley et al. (2001).

The simulation started with $195\,000$ single stars and $5\,000$ binaries.
We will refer to this as the K200 model. 
The binary fraction of 0.025 is guided by the findings of Davis et al. (2008) which
indicated a present day binary fraction of $\sim 0.02$ for 
the globular cluster NGC$\,6397$, measured near the half-light radius of the cluster. 
As shown in Hurley, Aarseth \& Shara (2007) and discussed in Hurley et al. (2008), 
this can be taken as representative of the initial binary fraction of the cluster. 
Thus we adopted this value for our model. 
Validation of the binary fraction approach will be provided in Section 7. 

Masses for the single stars were drawn from the initial mass function (IMF)
of Kroupa, Tout \& Gilmore (1993) between the mass limits of 0.1 and $50 M_\odot$.
Each binary mass was chosen from the IMF of Kroupa, Tout \& Gilmore (1991),
as this had not been corrected for the effect of binaries,
and the component masses were set by choosing a mass-ratio from a uniform distribution.
In {\tt NBODY4} we assume that all stars are on the zero-age main sequence when
the simulation begins and that any residual gas from the star formation
process has been removed.
A metallicity of $Z = 0.001$ was set for all stars.
The orbital separations of the $5\,000$ primordial binaries were drawn from the
log-normal distribution suggested by Eggleton, Fitchett \& Tout (1989) with a peak
at $30\,$au and a maximum of $100\,$au.
Orbital eccentricities of the primordial binaries were assumed to follow a
thermal distribution (Heggie 1975).

For the tidal field of the parent galaxy we have used the point-mass galaxy approach 
with the model cluster on a circular orbit at $R_{\rm gc} = 3.9\,$kpc with 
an orbital velocity of $V_{\rm g} = 220 \, {\rm km} \, {\rm s}^{-1}$.  
In setting $R_{\rm gc}$ we have been primarily guided by a desire for the cluster to 
have its moment
of core-collapse between $10 - 13\,$Gyr. 
Previous experience suggested that $R_{\rm gc} \simeq 4\,$kpc would provide 
this for a model starting with $N = 200\,000$. 

We used a Plummer density profile (Plummer 1911; Aarseth, H\'{e}non \& Wielen 1974) 
and assumed the stars
and binaries are in virial equilibrium when assigning the initial positions and
velocities.
The Plummer profile formally extends to infinite radius so in practice a cut-off at a radius
of $\sim 10 \, r_{\rm h}$ is applied, where $r_{\rm h}$ is the half-mass radius.
This is to avoid rare cases of large distance in the initial distribution. 
The tidal field sets a tidal radius 
according to:
\begin{equation}\label{e:rtide}
r_{\rm t} = \left( \frac{G M}{3}  \right)^{1/3} \, \left( \frac{V_{\rm g}}{R_{\rm gc}}  \right)^{2/3} \, ,
\end{equation}
where $G$ is the gravitational constant and $M$ is the cluster mass 
(see Giersz \& Heggie 1997). 
We 
chose the $N$-body length-scale of our model so that the outermost
star of the initial model sits at $r_{\rm t}$. 
This reflects the expansion expected when gas leftover from star formation 
is removed from the potential well (which we do not model), hence our initial 
model should be more radially extended than a compact protocluster. 
Stars were removed from the simulation when their distance from the density
centre exceeds twice that of the tidal radius of the cluster.

With these choices the initial parameters of the K200 model were
$M = 100\,067.4 \, M_\odot$, $r_{\rm h} = 4.7\,$pc and $r_{\rm t} = 35.8\,$pc.
The half-mass relaxation timescale of the initial model was $t_{\rm rh} \simeq 1\,115\,$Myr.

Much of the behaviour of this model will be compared to that reported
by Hurley et al. (2008) for a model that started with 
$95\,000$ single stars and $5\,000$ binaries. 
This will be referred to as the K100 model. 
The K100 model was placed on a circular orbit about a point-mass
galaxy at a radial distance of $R_{\rm gc} = 8.5\,$kpc. 
It had the same number of binaries as the K200 model but twice 
the binary fraction. 
The parameters of the stars and binaries were set up in the same 
manner as described above for the K200 model.

\begin{figure}
\includegraphics[width=84mm]{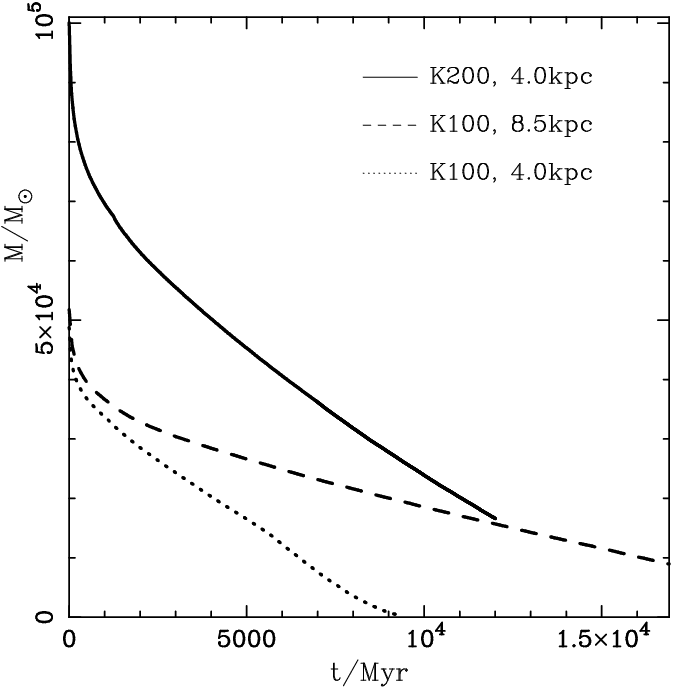}
\caption{
Evolution of the total cluster mass as a function of time for the
$N = 200\,000$ model orbiting at $R_{\rm gc} = 3.9\,$kpc (solid line),
the Hurley et al. (2008) $N = 100\,000$ (K100) model orbiting at $R_{\rm gc} = 8.5\,$kpc 
(dashed line)
and a $N = 100\,000$ model orbiting at $R_{\rm gc} = 4.0\,$kpc (dotted line).
\label{f:fig1}}
\end{figure}

\section{General Evolution}
\label{s:results}

In Figure~\ref{f:fig1} we look at the evolution of the total cluster mass with time
for the K200 model.
This simulation was stopped for analysis at $12\,$Gyr having satisfied the goal
of providing a post-core-collapse model (see below) at the approximate age
of a Milky Way globular cluster. 
At this point the model cluster has lost $83\%$ of its initial mass. 
In terms of stars remaining at $12\,$Gyr the K200 model has $N = 31\,713$
comprised of $30\,835$ single stars and $878$ binaries.

We include the evolution of the K100 model from Hurley et al. (2008) 
in Figure~\ref{f:fig1} and see that at $12\,$Gyr the two model clusters 
have the same amount of mass remaining (after starting with a factor 
of two difference). 
For comparison we also show in Figure~\ref{f:fig1} a model that started
with $100\,000$ stars on the same orbit as the K200 model.
As expected this model does not last for a Hubble time and is completely
dissolved after about $9\,$Gyr.
The slope in the mass-age plane is similar to that of the K200 model 
and distinct from that of the K100 model with $R_{\rm gc} = 8.5\,$kpc. 
An investigation of the mass-loss rates and dissolution times of
star clusters as a function of orbit within the Galaxy will be the subject
of another paper (Madrid et al. 2012). 

\begin{figure}
\includegraphics[width=84mm]{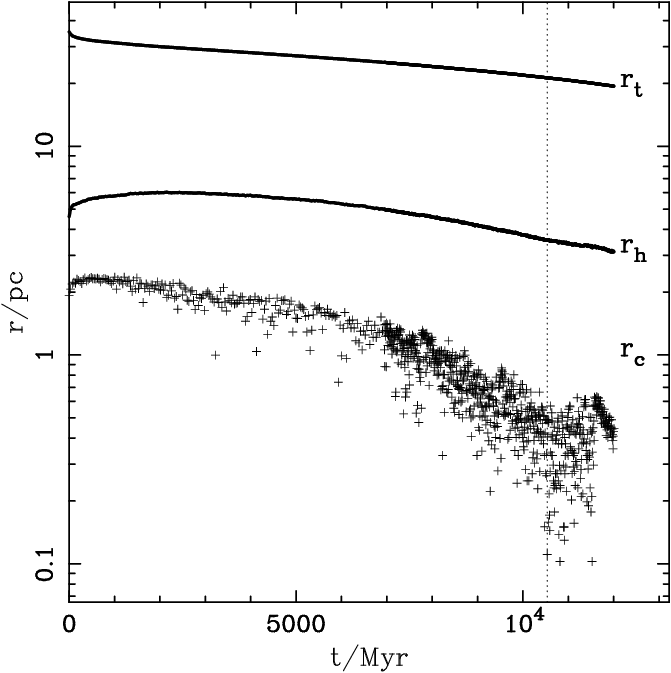}
\caption{
Evolution of the $N$-body core radius (+ symbols), the half-mass radius
(lower solid line) and tidal radius (upper solid line) for the
$N = 200\,000$ model.
All radii are three-dimensional. 
The vertical dotted line at $10.5\,$Gyr denotes the time we have identified with 
the end of the initial core-collapse phase. 
\label{f:fig2}}
\end{figure}

Figure~\ref{f:fig2} shows the behaviour of the core radius, the half-mass radius
and the tidal radius as the K200 model evolves.
Both the half-mass and core radii show an initial increase corresponding to
stellar evolution mass-loss from massive stars, 
which are mostly found in the inner regions of the cluster. 
The half-mass radius then plateaus before appearing to follow the
decreasing trend of the tidal radius at later times. 
At $12\,$Gyr we have $r_{\rm h} \simeq 3\,$pc which is comparable 
to the average effective radius found for globular clusters 
(e.g. Jord\'{a}n et al. 2005). 

The core-radius shows a deep minimum at $10.5\,$Gyr which we
identify as the moment that the initial core-collapse phase ends. 
This is based purely on inspection of Figure~\ref{f:fig2}, noting that the 
same method 
-- looking for the first deep minimum of the density- or mass-dependent 
central radius 
-- has been commonly employed in the past (e.g. 
Baumgardt \& Makino 2003; Hurley et al. 2004; K\"{u}pper et al. 2008). 
At this point the core density is $5\,200\, {\rm stars} \, {\rm pc}^{-3}$,
increased from an initial value of $700 \, {\rm stars} \, {\rm pc}^{-3}$. 
Subsequently the core fluctuates markedly, corresponding to 
post-core-collapse oscillations highlighted by Heggie \& Giersz (2009). 
However, we note that the core exhibits fluctuating behaviour leading up 
to the moment of core-collapse as well. 

The core radius in Figure~\ref{f:fig2} is the density radius commonly
used in $N$-body simulations (Casertano \& Hut 1985), 
calculated from the density weighted average of the distance of each
star from the density centre (Aarseth 2003). 
Following Heggie \& Giersz (2009) we have also looked at the dynamical  
core radius used in their Monte Carlo models, calculated as 
$r_{\rm c,D}^2 = 3 \, v_{20}^2 / \left( 4 \pi G \, \rho_{20} \right)$ 
where $v_{20}$ is the mass-weighted central velocity dispersion and $\rho_{20}$
is the central density, both calculated from the innermost 20 stars. 
We find that $r_{\rm c}$ and $r_{\rm c,D}$ cover a similar range at all times. 
In particular, for the period $10.5 - 11.5\,$Gyr, i.e. $1\,$Gyr after core-collapse, 
both oscillate between 0.1 to $0.6\,$pc for the majority of the time. 
For reference, the radius containing the inner 1\% of the cluster mass 
is $\sim 0.3\,$pc over this period and is thus a good proxy for the average 
core radius. 
Also following Heggie \& Giersz (2009) we have used the autocorrelation method to
determine if there is any clear periodicity in the fluctuations of the $N$-body core radius 
in the $1\,$Gyr subsequent to core-collapse. 
This showed a period of about $120\,$Myr: greater than the crossing time (few Myr) 
and less than the relaxation time ($300 - 400\,$Myr). 
In comparison, Heggie \& Giersz (2009) reported an oscillation period of 
$\sim 400\,$Myr for their $N$-body model\footnote{
This $N$-body model was evolved for $1\,$Gyr starting from a post-collapse 
Monte Carlo model at an age of $12\,$Gyr.
}, 
although this contained more $N$ 
than our post-collapse K200 model and correspondingly had a higher 
half-mass relaxation timescale of $\sim 700\,$Myr. 

It is of interest to examine how previous $N$-body results reported for smaller $N$ models hold
up in comparison to our new model. 
The K100 model of Hurley et al. (2008) reached a similar core radius at the end 
of core-collapse as did our K200 model ($r_{\rm c} \sim 0.1\,$pc), albeit at a 
later time ($16\,$Gyr compared to $10.5\,$Gyr). 
As noted above, the average value of $r_{\rm c}$ near core-collapse is similar to 
the 1\% Lagrangian radius for the K200 model. 
For the K100 model $r_{\rm c}$ evolves similarly to the 2\% Lagrangian radius 
(although it does dip down to the 1\% radius on occasion) 
and for the $N = 500$ models of Giersz \& Heggie (1997) $r_{\rm c}$ is at the 
5\% Lagrangian radius or greater. 
Thus it seems that with increasing $N$ the depth of core-collapse increases relative to the 
cluster mass distribution. 
If we look instead at scaled quantities, particularly the evolution of the 
ratio $r_{\rm c} / r_{\rm h}$ as a function of age scaled by the half-mass relaxation timescale, 
we find that the K200 and K100 simulations track each other very well. 
The ratio starts at $\sim 0.4$ and steadily decreases to an average value of $\sim 0.1$ at 
the end of core-collapse. 
Giersz \& Heggie (1997) find $r_{\rm c} / r_{\rm h} \sim 0.15$ for their models 
at core-collapse, noting that they use $r_{\rm c,D}$ rather than the 
Casertano \& Hut (1985) definition and that latter is about twice as large in 
the post-collapse phase for their models (we found that the two gave similar 
average values). 
McMillan (1993) performed models with $N = 1\,024$, primordial binaries and 
a tidal field. 
For these models $r_{\rm c} / r_{\rm h}$ stabilized at about 0.1 in good agreement 
with prior models of isolated clusters (see McMillan 1993 for details). 
It therefore seems that this can be taken as a reliable value for clusters at 
the end of core-collapse (although see the next section for mention of some 
unusual cases).

\section{The late action of a black-hole binary}
\label{s:bhbinary}

A remarkable feature in Figure~\ref{f:fig2} is the sharp change in the behaviour of the 
core at $\sim 11.5\,$Gyr, the last deep minimum, when the size of the 
core suddenly increases and evolves steadily from that point onwards. 
This change is related to an interaction within the core involving a binary comprised 
of two BHs. 

The binary in question is non-primordial. 
Each BH formed from a massive single star within the first $10\,$Myr of evolution 
with masses of $23.6$ and $17.3 \, M_\odot$, respectively. 
The two BHs formed a binary at $1\,800\,$Myr in a four-body interaction, 
initially with a very high eccentricity and long orbital period of $10^5\,$d.
It resided in the core for the majority of its lifetime and suffered a series 
of perturbations and interactions that saw the eccentricity vary between 0.2 to 0.95 
and the orbital period reduced to $13\,$d at $11\,550\,$Myr. 
At that time the BH-binary becomes embroiled in a strong interaction with 
a binary comprised of two main-sequence stars (masses of $0.8$ and $0.6 \, M_\odot$). 
The second binary is broken-up and the two main-sequence stars are ejected 
rapidly from the cluster (velocities of $96$ and $154 \, {\rm km} \, {\rm s}^{-1}$). 
This causes a recoil of the BH-binary which leaves the core and then the 
cluster entirely ($10\,$Myr later with a velocity of $5 \, {\rm km} \, {\rm s}^{-1}$). 
The domination of the central region of the cluster by this BH-binary and its 
subsequent ejection are similar to the processes described by Aarseth (2012). 

The sudden loss of mass from the core -- the average mass drops by 30\% 
(see next section) -- combined 
with the rapid ejection of the two main-sequence stars causes the core to expand. 
We see that after this event the core radius does start to decrease once more but 
without fluctuations. 
Thus, the influence of one strong interaction involving a massive binary has 
halted the core oscillation process. 
Compared to the point that we identified as the end of the initial core-collapse 
phase the core radius has increased by a factor of about six. 
The structure of globular clusters is often quantified by the concentration 
parameter $c = \log r_{\rm t} / r_{\rm c}$ (King 1966). 
Milky Way GCs exhibit a range of $c$ values (Harris 1996) with the most obvious 
core collapse examples having $c = 2.5$ but with $c > 2$ generally taken as 
indicative of a high-density cluster or a possible core-collapse cluster (Mateo 1987). 
At the end of the core-collapse phase our K200 model has $c \sim 2.3$ and 
this decreases to $c \sim 1.5$ after the BH-binary is ejected from the core. 
Thus, the cluster would not be expected to appear as a core-collapse cluster 
if observed at this point. 
 
Hurley (2007) showed that the presence of a long-lived BH-BH binary in the core, with
both BHs being of stellar mass, could significantly increase the $r_{\rm c} / r_{\rm h}$ ratio
of a model with $100\,000$ stars. 
The BH-BH binary in our $N = 200\,000$ model has not produced a similar 
inflation of the ratio. 
Mackey et al. (2007) performed $N$-body simulations with $N \sim 100\,000$ in which 
they retained $\sim 200$ stellar mass BHs. 
They found that the BHs formed a dense core in which interactions were common and 
BHs could be ejected from the cluster, leading to a significantly inflated core radius. 
It is our intention in the near future to look at a wide range of $N$-body simulations 
and document in detail the statistics and outcomes of BH-BH binaries in the cores 
of model clusters. 
This will include fitting King (1966) models to the density profiles of the model clusters 
so as to properly calculate the concentration parameter rather than using $N$-body 
values as we have done here in our preliminary analysis.

\begin{figure}
\includegraphics[width=84mm]{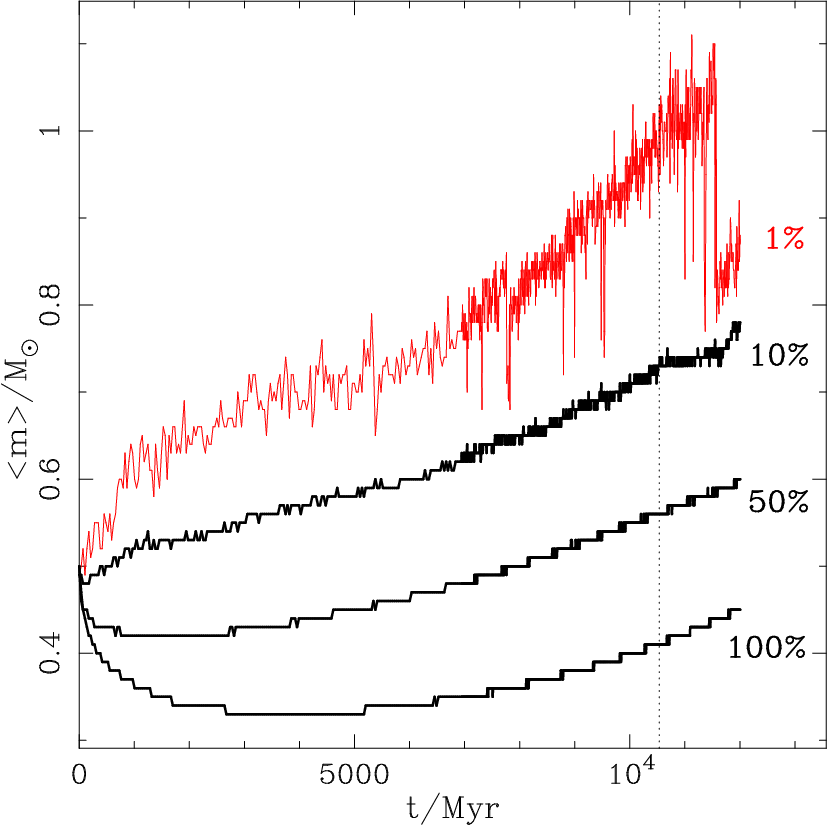}
\caption{
Average mass in different Lagrangian regions (0-1\%, 1-10\%, 10-50\%, 50-100\%) for the
$N = 200\,000$ model. 
The vertical dotted line marks the end of the initial core-collapse phase 
as in Figure~\ref{f:fig2}. 
\label{f:fig3}}
\end{figure}

\section{Mass segregation}
\label{s:mseg}

Figure~\ref{f:fig3} looks at how the average stellar mass behaves for
the K200 model, focussing on four different Lagrangian regions:
a central volume that encompasses the inner 1\% of the cluster mass,
a central shell that lies between radii enclosing 1\% and 10\% of the
cluster mass,
an intermediate shell that lies between the 10\% and 50\% Lagrangian radii,
and an outer shell that includes all stars beyond the 50\% Lagrangian radius.
Note that binaries are included and are assumed to be unresolved.
The average stellar mass throughout the entire cluster is $0.5 M_\odot$ at
the start of the simulation
-- there is no primordial mass segregation -- and
drops initially in all regions owing to stellar evolution mass-loss of
massive stars for the first $\sim 100\,$Myr.
We then see that the effect of mass segregation driven by two-body encounters
takes over, causing an increase in the average stellar mass in the inner regions and
a corresponding decrease in the outer region.
By the time that one half-mass relaxation timescale has elapsed ($\sim 1\,000\,$Myr)
there is a clear distinction between the average mass in each of the regions.
In the central regions the average stellar mass continues to increase up to the
end of core-collapse and then flattens post-collapse 
(until the ejection of the BH-BH binary). 
The value in the very centre is noisy owing to a smaller number of objects and
the cycling of these objects in and out of the region.
We see a marked increase in this central value as the cluster gets closer to
the end of core-collapse (note the correlation with $r_{\rm c}$ in Figure~\ref{f:fig2}).
The decreasing average mass in the outer region is gradually arrested by the
effect of the external tide which preferentially removes the lower-mass stars that
have been pushed out to this region.
We see that from $\sim 5\,000\,$Myr onwards the average mass in the outer
region is now increasing and that this effect is also felt in the intermediate region.

\begin{figure}
\includegraphics[width=84mm]{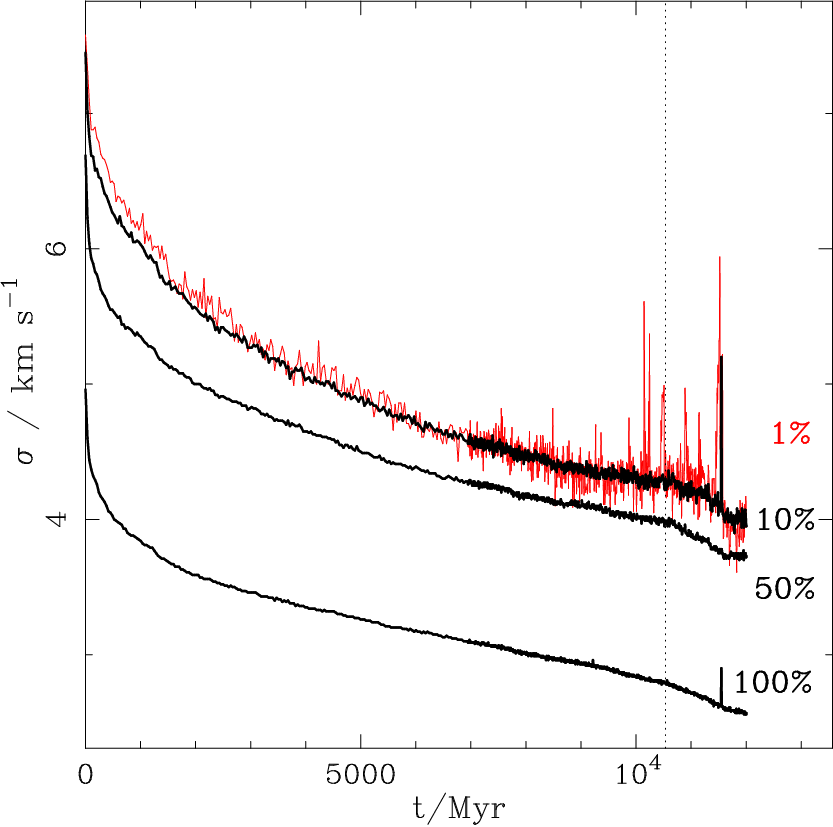}
\caption{
As for Figure~\ref{f:fig3} but now showing the velocity dispersion.
\label{f:fig4}}
\end{figure}

If we instead look at the behaviour in two-dimensional projected regions we find that 
the average mass is the same in the two outermost regions but drops by about 5\% 
in the 1-10\% region and 10-15\% in the central region (after the first Gyr). 

Giersz \& Heggie (1997) look at the evolution of average mass in different regions, as
we have done in Figure~\ref{f:fig3}. 
They see very similar behaviour which can be summarised as:  
(i) a sharp increase of the average mass within the 1\% Lagrangian region towards core-collapse; 
(ii) a decrease in the outer regions that flattens out with time and then increases at late times; 
and (iii) similar trends in the intermediate regions. 
However, in their $N = 500$ models they find that the timescale for the initial phase 
of mass segregation is about the same as the core-collapse timescale, 
whereas we find that mass segregation is fully established well before core-collapse. 
The $N = 131\,072$ models of Baumgardt \& Makino (2003) agree with our K200 model 
in that respect and with the general behaviour, including the flattening of the average 
mass in the central regions post-collapse.

\begin{figure}
\includegraphics[width=84mm]{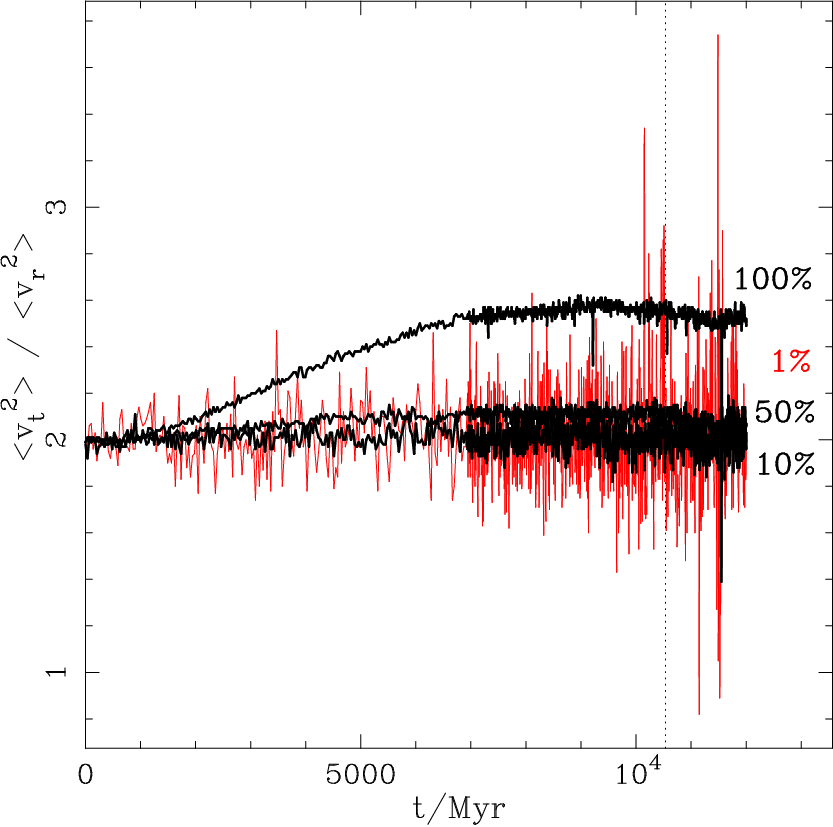}
\caption{
As for Figure~\ref{f:fig3} but now showing the anisotropy parameter. 
\label{f:fig5}}
\end{figure}

\section{Velocity distributions}
\label{s:velocity}

In Figure~\ref{f:fig4} we show the velocity dispersion for the
same Lagrangian regions as in Figure~\ref{f:fig3}.
All regions show a rapid initial decrease owing to an overall expansion of the cluster.
This is followed by a more gradual decrease as the cluster evolves towards
core-collapse, with all regions declining in an almost homologous manner.
As the model cluster nears the end of core-collapse there is a pronounced upturn in the
velocity dispersion within the 1\% radius.
This is also seen out to the 10\% radius and even at the half-mass radius, although
to a much smaller extent. 
Note that the behaviour for two-dimensional velocities in projected Lagrangian 
regions is the same but with values that are typically 20\% less. 

The main features of our new model mirror those in the 
smaller-$N$ models of Giersz \& Heggie (1997). 
These features are the following: 
the values within the 1\% and 10\% regions overlap; 
the velocity dispersion in the outer regions is clearly lower than for these inner regions; 
and the values for the 50\% region are much closer to those of the inner regions than
the outer regions.

Another aspect of velocity to consider is the anisotropy. 
We have chosen to define this as the ratio of the mean square transverse 
to radial velocity components, 
$<v_{\rm t}^2> / <v_{\rm r}^2>$ (as in Giersz \& Heggie 1997) 
and the result is shown in Figure~\ref{f:fig5}. 
In three dimensions this is equal to two for isotropy, greater than two for a 
tangentially anisotropic distribution and less than two for a radially anisotropic 
distribution. 
An alternative is to use the anisotropy parameter $\beta = 1 - <v_{\rm t}^2> / 2  <v_{\rm r}^2>$ 
(Binney \& Tremaine 1987; Baumgardt \& Makino 2003; Wilkinson et al. 2003) 
where $\beta = 0$ is isotropic, $\beta < 0$ is tangentially anisotropic 
and $\beta > 0$ is radially anisotropic. 
We see from Figure~\ref{f:fig5} that the inner regions of the cluster remain close 
to isotropic throughout the evolution while the outer region develops an increasing 
tangential anisotropy over the first $\sim 7\,$Gyr and then flattens out to a 
constant value for the remainder of the evolution. 
This overall behaviour is similar to that observed by Baumgardt \& Makino (2003) in 
their models. 
It is also similar to the behaviour for the K100 model, although the degree of anisotropy in 
the outer region is about 30\% less than in the K200 model. 
Also, because the K100 model is evolved well past core-collapse it is possible to see that 
the anisotropy is reduced post-collapse and tends towards isotropy in the final stages 
of evolution (as was noted by Baumgardt \& Makino 2003). 
The tangential anisotropy in the outer region is contrary to previous findings 
of radial anisotropy for smaller-$N$ models (Giersz \& Heggie 1997; Wilkinson et al. 2003). 
Tangential anisotropy can be explained by stars expelled from the central regions on radial 
orbits preferentially escaping from the cluster at the expense of stars on tangential orbits 
which find it harder to escape. 
In the very central region (within the 1\% Lagrangian radius) the anisotropy parameter 
is very noisy and fluctuates between radial and tangential anisotropy. 
However, the average behaviour is isotropy.

\begin{figure}
\includegraphics[width=84mm]{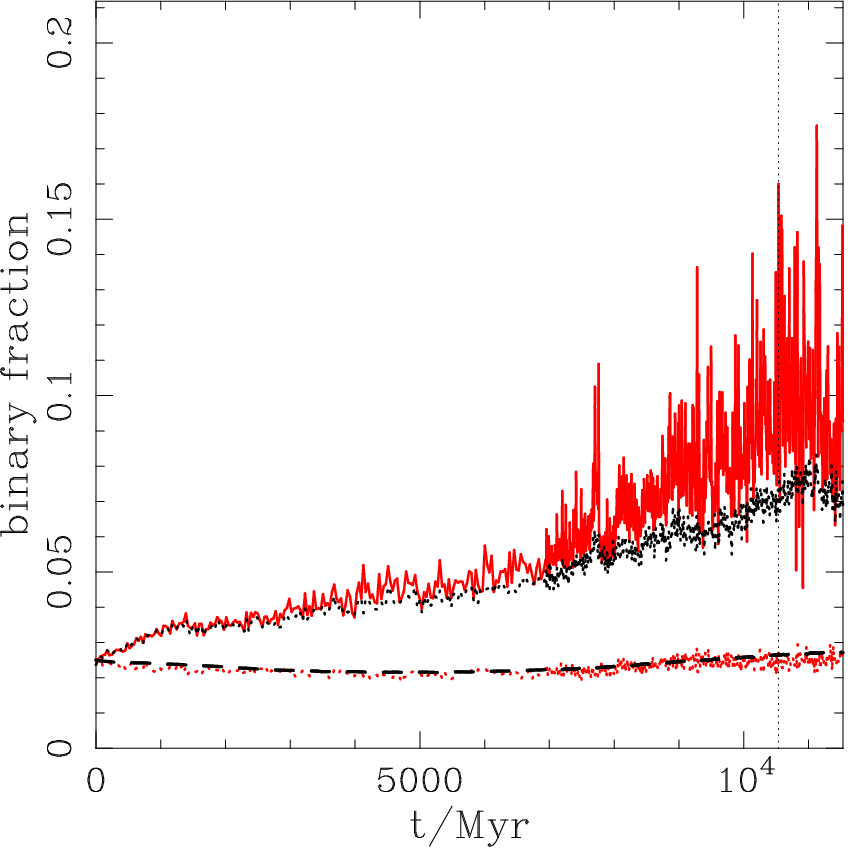}
\caption{
Evolution of the binary fraction for the $N = 200\,000$ model.
Shown are the binary fraction of the entire cluster (dashed line),
within the 10\% Lagrangian radius (black dotted line) and within
the core (red solid line).
Also shown is the binary fraction of main-sequence stars and binaries near the
half-mass radius (red dotted line: which closely follows the 
binary fraction of the entire cluster).
Compare with Figure~3 of Hurley, Aarseth \& Shara (2007). 
The vertical dotted line once again marks the end of the initial core-collapse phase 
as in Figure~\ref{f:fig2}. 
\label{f:fig6}}
\end{figure}

\section{Binaries}
\label{s:ebinary}

Hurley, Aarseth \& Shara (2007) documented the evolution of binary fraction with time
for a range of $N$-body models covering $N = 24\,000$ to $100\,000$ and
primordial binary fractions from 0.05 to 0.5.
In all cases they found that the binary fraction within the cluster core and the
10\% Lagrangian radius increases as the cluster evolves towards core-collapse
while the overall binary fraction of the cluster stays close to the primordial value. 
Figure~\ref{f:fig6} shows the same evolution for the K200 model. 
Similar results are seen in that the core binary fraction increases markedly as
the cluster evolves, particularly towards core-collapse when it becomes a
factor of six or more greater than the primordial value, while the overall binary
fraction does stay close to the primordial value (although the inclusion of a large
proportion of initially wide binaries would lead to an initial decrease). 
This gives added confidence in the results of Hurley, Aarseth \& Shara (2007)
although models of greater $N$ (and density) are still required before the
situation for the majority of the Milky Way globular clusters can be firmly established.

Davis et al. (2008) measured the binary fraction amongst main-sequence
stars near the half-mass radius of NGC$\,6397$ and found it to be a few per cent.
This was taken as representative of the primordial binary fraction of NGC$\,6397$
by leaning on the result from Hurley et al. (2008) showing that the
binary fraction measured near the half-mass
radius of a cluster can be taken as a good indication of the primordial binary fraction.
This is reinforced in Figure~\ref{f:fig6} for our K200 model where we show that 
the binary fraction of main-sequence stars and binaries near the half-mass radius 
of the cluster stays at roughly the same value throughout the evolution. 

\begin{figure}
\includegraphics[width=84mm]{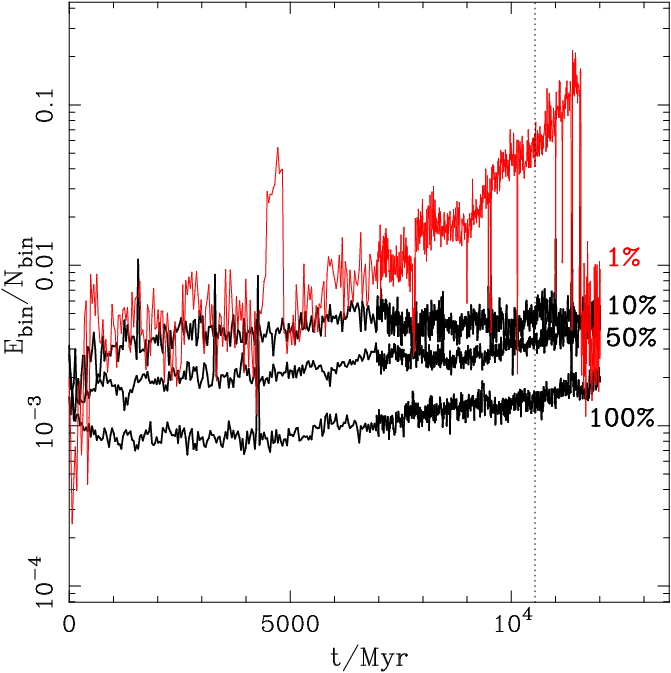}
\caption{
Average energy per binary in different Lagrangian regions 
(0-1\%, 1-10\%, 10-50\%, 50-100\%) for the
$N = 200\,000$ model. 
The vertical dotted line denotes the time identified with 
the end of the initial core-collapse phase for this model. 
\label{f:fig7}}
\end{figure}

\begin{figure}
\includegraphics[width=84mm]{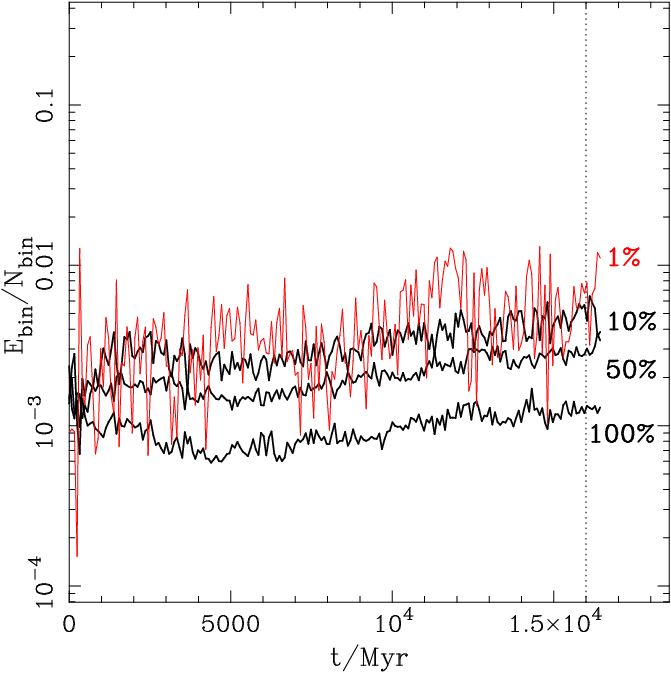}
\caption{
As for Figure~\ref{f:fig7} but now for the K100 model. 
The vertical dotted line denotes the time identified with 
the end of the initial core-collapse phase for this model. 
\label{f:fig8}}
\end{figure}

Next we look at the energy in binaries for the K200 simulation in Figure~\ref{f:fig7} 
and for the K100 simulation in Figure~\ref{f:fig8}. 
Here we see that for the $10 - 100$\% Lagrangian mass regions the 
values and behaviour for the two simulations are very similar. 
As  expected the average energy per binary increases as
we move inwards towards the cluster centre, although the values within the
1-10\% and 10-50\% regions are converging towards the end of the simulation. 
We note that the binary binding energy is calculated in arbitrary yet physical 
units of $M_\odot^2 / R_\odot$ to enable direct comparison between the 
internal energies of the two binary populations. 
Conversion to cluster units of $k$T based on the average kinetic energy of the 
cluster stars (e.g. McMillan 1993) increases the K100 values by a factor of three compared to the 
K200 values, i.e. the binaries have relatively more internal energy in the 
simulation with less stars. 
Sharp dips in the binding energy occur throughout the simulations when 
binaries escape (as documented by K\"{u}pper, Kroupa \& Baumgardt 2008) 
although these are less evident in Figure~\ref{f:fig8} owing to less frequent sampling. 

The energy per binary in the inner 1\% region by mass starts off similarly for 
both simulations, oscillating about the 1-10\% values. 
However, the influence of the BH-binary in the K200 simulation from $\sim 5\,$Gyr 
onwards is clear to see, increasing the average energy by almost two orders 
of magnitude before a sharp drop when the binary is removed from the core. 

Noticeable increases in binding energy 
are also evident at other times and arise from the creation of tight binaries
by various means.
For example, the spike in the 50-100\% Lagrangian region at $\sim 4\,300\,$Myr 
in Figure~\ref{f:fig7} results
from a primordial binary in which common-envelope evolution creates
a short-period binary comprised of two white dwarfs which subsequently come into contact
and merge.
Thus the binary dominates the energy in this region for a short time and then
disappears.
We see a more persistent increase in the average energy within the 1\% region
at $\sim 4\,500\,$Myr in Figure~\ref{f:fig7}.
This is caused when a main-sequence star and a black hole form a short-period
binary via an exchange interaction.
The binary survives for about $400\,$Myr in the cluster core before the main-sequence
star is consumed by the black hole.

It is worthwhile to ask if the signatures of these short-lived energetic binaries be observed?
The binary that resulted in the merger of two WDs did not produce a WD with a mass
in excess of the Chandrasekhar limit so could not be a potential type Ia supernova
(see Shara \& Hurley 2002).
However, the resulting WD will be relatively massive, hot and {\it young}.
Thus it could be expected to remain one of the brightest WDs in the cluster
for several Gyrs and would be easily observed.
The other binary mentioned resulted in a main-sequence star being tidally disrupted and
swallowed by a black-hole of $\sim 20 \, M_\odot$.
This could be expected to give rise to a burst of X-rays and perhaps gamma-rays,
along the lines of Burrows et al. (2011: albeit for a supermassive black hole).

\begin{table}
\begin{minipage}{80mm}
\caption{
Comparison of core-collapse, $t_{\rm cc}$, and relaxation timescales, $t_{\rm rh}$,
for the K200 and K100 simulations.
Note that the averaged $t_{\rm rh}$ is calculated to $11.5\,$Gyr and $18\,$Gyr for
the K200 and K100 simulations, respectively.
Also shown is the time for the cluster to lose half of the initial mass, $t_{1/2}$.
\label{t:table1}
}
\begin{tabular}{lrr}
\hline  
quantity & K200 & K100 \\
\hline
$t_{1/2}$/Myr & 4042 & 5415 \\
$t_{\rm cc}$/Myr & 10500 & 16000 \\
$t_{\rm rh}$/Myr (initial) & 1115 & 1405 \\
$t_{\rm rh}$/Myr ($t_{1/2}$) & 1480 & 1910 \\
$t_{\rm rh}$/Myr (cc) & 415 & 500 \\
$t_{\rm rh}$/Myr (average) & 762 & 1340 \\
\hline
\end{tabular}
\end{minipage}
\end{table}

\section{Scaling of previous timescale results}

Timescales for star cluster evolution are important to understand, with the time until
core-collapse and the time until dissolution being quantities of interest 
(e.g. Gnedin \& Ostriker 1997).
Furthermore, the scaling of these with $N$ or related properties of clusters/models is
necessary (Baumgardt 2001),
particularly while direct models of globular clusters remain out of reach.

In Table~\ref{t:table1} we summarize the significant timescales for the
K200 and K100 simulations: the time for half of the initial mass to be lost, 
the time until the end of the core-collapse phase and the 
half-mass relaxation time at various key points. 

K\"{u}pper et al. (2008) presented a range of open cluster models in a steady
tidal field and found that the core-collapse time scaled by the initial half-mass relaxation time
ranged from 17 for clusters starting well within their tidal radii (smaller $t_{\rm rh,0}$)
to 9 for clusters that fill their tidal radii from the start (larger $t_{\rm rh,0}$).
Our K200 model and the K100 model of Hurley et al. (2008) start by filling their tidal 
radius and have $t_{\rm cc} / t_{\rm rh,0} \sim 10$
which is in good agreement with K\"{u}pper et al. (2008).

Baumgardt (2001) looked at the scaling of $N$-body models using simulations
of up to $16\,384$ equal-mass stars.
In this work the time for a cluster to lose half of its mass, $t_{1/2}$, was taken as an
indication of the cluster lifetime with $t_{1/2} \propto t_{\rm rh}^{3/4}$ shown to be 
the appropriate scaling. 
Baumgardt \& Makino (2003) subsequently used their larger-$N$ models 
(up to $131\,072$ stars) to once again show
that scaling the dissolution time as $t_{\rm rh}^{3/4}$ is appropriate, this time for
multi-mass models that included stellar evolution. 
However, we see from Table~\ref{t:table1} that the half-mass relaxation time varies
considerably across the lifetime of a cluster and it is not immediately clear which value
to use when scaling timescales.
For an observed cluster it will be the value at the current age of the cluster. 
A fairer comparison would be to use the average half-mass relaxation time
across the lifetime of the cluster, $t_{\rm rh,av}$
-- of course this is not known for an observed cluster.
For the K200 and K100 models we find that $t_{\rm cc} / t_{\rm rh,av}^{3/4}$
is the same for both simulations, so the agreement for this key timescale is
in excellent agreement with the previous suggestion.
We also find that $t_{1/2}$ scales quite well with $t_{\rm rh,av}^{3/4}$.
However, if we instead use $t_{\rm rh,0}^{3/4}$ then we do not find good agreement
for either the core-collapse or half-life timescales.
Thus our agreement with the scalings found in previous works is dependent
on which $t_{\rm rh}$ is used.

\section{Summary}
\label{s:discus}

We have presented an $N$-body model that started with $200\,000$ stars and binaries, 
evolves to the moment of core-collapse at $10.5\,$Gyr and has $\sim 30\,000$ stars 
remaining at $12\,$Gyr. 
We have used our direct $N$-body model  
to confirm the post-core-collapse fluctuations described in the Monte Carlo 
model of Heggie \& Giersz (2008) and the hybrid $N$-body/MC approach 
of Heggie \& Giersz (2009). 
We have also shown that these fluctuations can be halted by the ejection 
of a dominant BH-binary from the core. 
This produces a core that shows no sign that it has previously evolved 
through core-collapse. 

We have looked at how the results of previous works compare to a model 
of larger $N$ and find good agreement provided that appropriate scalings 
are used (such as the core-radius to half-mass radius ratio at core-collapse). 
In terms of raw values some variations exist: the core radius at core-collapse 
reaches deeper in to the mass distribution for larger $N$, for example. 
The behaviour of quantities such as average stellar mass and velocity dispersion have been
documented and the general behaviour matches expectations from earlier models.
Looking at time scales such as the time to core-collapse 
and the dissolution time we also find agreement with scaling relations 
previously reported in the literature, however this is 
dependent on which value of the half-mass relaxation timescale is used. 
In particular, the scaling of dissolution time with $t_{\rm rh}^{3/4}$ reported 
by Baumgardt (2001) could be reproduced provided that the average $t_{\rm rh}$ 
was used and not the initial $t_{\rm rh}$.

The $N = 200\,000$ simulation reported in this work took the best part of a year
on a GRAPE-6 board to complete.
It continues the gradual increase of $N$ used in realistic $N$-body models from
the $N = 500$ model of Giersz \& Heggie (1997)
to the $N = 131\,072$ model of Baumgardt \& Makino (2003).
However, with only $\sim 20\,000 \, M_\odot$ remaining in our model at an
age of $11.5\,$Gyr we are still only touching the lower end of
the globular cluster mass-function.
This is after considerable effort.
In particular, we are still some way from the goal of a full million-body model of
a globular cluster (Heggie \& Hut 2003).
How can we push forward to reach that goal?
The shift towards graphics processing units (GPUs) as the central computing engine 
for $N$-body codes,
combined with sophisticated software development, offers hope (Nitadori \& Aarseth 2012).
Simulations of $100\,000$ stars can be performed comfortably on a single-GPU
(Hurley \& Mackey 2010; Zonoozi et al. 2011)
and the introduction of multiple-GPU support will likely make simulations of the type
presented here commonplace in the near future.
Further hardware advances and a revisiting of efforts to parallelize direct 
$N$-body codes (Spurzem 1999) will also aid the push towards greater $N$.

In a follow-up paper we will conduct a full investigation of the stellar and 
binary populations of our model. 
It is also our intention to make model snapshots 
-- saved at frequent intervals across the lifetime of the simulation -- available 
for others to {\it observe} and analyse. 
These can be obtained by contacting the authors. 
At the beginning of this paper we indicated that an important function of a large-$N$
model would be to aid in the calibration of the Monte Carlo technique.
This is currently underway (Giersz et al. 2012) and will
include a direct comparison of $N$-body and MC models starting from the same
initial conditions.

\section*{Acknowledgments}
We acknowledge the generous support of the Cordelia Corporation and
that of Edward Norton which has enabled AMNH to purchase
GRAPE-6 boards and supporting hardware. 
We thank Harvey Richer for helping to provide the motivation for this model 
and Ivan King for many helpful suggestions. 


\label{lastpage}

\end{document}